\documentclass[12pt]{article}
\usepackage{graphicx}
\usepackage{amssymb}
\usepackage{amsmath}
\newcommand{\vev}[1]{{\langle #1 \rangle}}
\newcommand{\abs}[1]{{\left| #1 \right|}}

\newcommand{\eV}{\mbox{~eV}}

\newcommand{\GeV}{\mbox{~GeV}}
\newcommand{\TeV}{\mbox{~TeV}}
\newcommand{\ie}{{\it i.e.}}
\newcommand{\eg}{{\it e.g.}}
\newcommand{\eqn}[1]{&\hspace{-0.6em}#1\hspace{-0.6em}&}
\begin{document}
\baselineskip 0.6cm
%
\begin{titlepage}
\begin{center}

\begin{flushright}
\end{flushright}

\vskip 2cm

{\Large \bf 
Flavour Mixing of Neutrinos and 
Baryon Asymmetry of the Universe
}

\vskip 1.2cm

{\large 
Takehiko Asaka$^1$ and Hiroyuki Ishida$^2$
}

\vskip 0.4cm

$^1${\em
  Department of Physics, Niigata University, Niigata 950-2181, Japan
}

$^2${\em
  Graduate School of Science and Technology, Niigata University, Niigata 950-2181, Japan
}

\vskip 0.2cm

(April 30, 2010)

\vskip 2cm

\vskip .5in
\begin{abstract}
We investigate baryogenesis in the $\nu$MSM, which is the Minimal
Standard Model (MSM) extended by three right-handed neutrinos with
Majorana masses smaller than the weak scale.  In this model the baryon
asymmetry of the universe (BAU) is generated via flavour oscillation
between right-handed neutrinos.  We consider the case when BAU is
solely originated from the CP violation in the mixing matrix of active
neutrinos.  We perform analytical and numerical estimations of
the yield of BAU, and show how BAU depends on mixing angles and CP
violating phases.  It is found that the asymmetry in the inverted
hierarchy for neutrino masses receives a suppression factor of about
4\% comparing with the normal hierarchy case.  It is, however, pointed
out that, when $\theta_{13}=0$ and $\theta_{23} = \pi/4$, baryogenesis
in the normal hierarchy becomes ineffective, and
hence the inverted hierarchy case becomes significant to account for
the present BAU.
\end{abstract}
\end{center}
\end{titlepage}
\renewcommand{\thefootnote}{\#\arabic{footnote}} 
\setcounter{footnote}{0}
%
%
\section{Introduction}
The origin of the baryon asymmetry of the universe (BAU) is one of the
most mysterious problems in particle physics and cosmology, since the
Minimal Standard Model (MSM) and the Big Bang cosmology cannot answer
it.  So far various mechanisms for generating BAU have been
proposed~\cite{Riotto:1999yt}.  One promising possibility is the
so-called leptogenesis scenario~\cite{Fukugita:1986hr} (see also
Ref.~\cite{Buchmuller:2005eh}), where the origins of neutrino masses
as well as BAU can be addressed at the same time by introducing
right-handed neutrinos with superheavy Majorana masses.  The observed
tiny masses of neutrinos can be naturally understood by the seesaw
mechanism~\cite{Seesaw}.  Further, the lepton asymmetry generated by
decays of right-handed neutrinos can be a source of BAU.  In the
simplest thermal leptogenesis, the required Majorana masses is heavier
than about $10^9$ GeV~\cite{Giudice:2003jh}.

It should be, however, noted that the connection between the origins
of neutrino masses and BAU can be obtained even when Majorana masses
are below the weak scale.  One interesting possibility is the
so-called $\nu$MSM~\cite{Asaka:2005pn,Asaka:2005an}, which is the
MSM extended by three right-handed neutrinos
with masses smaller than the weak scale.  In this model the problems
of neutrino masses, BAU and also dark matter can be solved
simultaneously.  One attractive advantage of the $\nu$MSM lies
in the fact that the direct detection of right-handed 
neutrinos becomes possible~\cite{Gorbunov:2007ak}.

In the $\nu$MSM BAU can be generated by invoking the mechanism via
flavour oscillation of right-handed neutrinos~\cite{Akhmedov:1998qx}.
(See also Ref.~\cite{Asaka:2005pn,Shaposhnikov:2008pf}.)  In this
mechanism the lepton asymmetry is not generated for temperatures of
interest because of the smallness of Majorana masses, which is very
different from the leptogenesis scenario.  The lepton asymmetry is
separated into left-handed and right-handed leptonic sectors due to
the CP violations in the production and oscillation of right-handed
neutrinos.  Then, the asymmetry stored in the left-handed sector is
partially transferred into the baryon asymmetry through the rapid
sphaleron transitions~\cite{Kuzmin:1985mm}.  

One of right-handed neutrinos in the $\nu$MSM, which is a candidate of
dark matter, plays no essential role to generate BAU as well as
neutrino masses observed in the oscillation experiments, since its
Yukawa coupling constants should be very suppressed.  The rest two are
responsible to BAU via their flavour oscillation, but also induce the
masses of active neutrinos through the seesaw mechanism.
Therefore, physics of these two right-handed neutrinos
connects BAU with the neutrino parameters of active neutrinos,
\ie, mass hierarchy, mixing angles, and CP violating phases.

In this letter we would like to extend the analysis in
Ref.~\cite{Asaka:2005pn}.  Under the considering situation
there are three CP violating phases in the
leptonic sector which can be a source of the asymmetry.  Especially,
we concentrate here on the case when BAU is originated only from the
CP violation in the mixing matrix $U$ of active neutrinos, namely, the
Dirac phase $\delta$ and Majorana phase $\eta$ in $U$.  We then
present the analytical expression of BAU shows explicitly how BAU
depends on these CP phases and the mixing angles of active neutrinos.
Moreover, we also perform the numerical estimation of BAU
and justify the validity of the analytical expression.
\section{The $\nu$MSM}
We begin with the brief review of the model under consideration, \ie,
the $\nu$MSM~\cite{Asaka:2005pn,Asaka:2005an}.  It is the MSM extended
by three right-handed neutrinos $\nu_{R \, I}$ ($I=1,2,3$), which
Lagrangian is given by
\begin{eqnarray}
  \label{eq:L_nuMSM}
  {\cal L}_{\nu{\rm MSM}} =
  {\cal L}_{\rm MSM} +
  i \, \overline{\nu_R{}_I} \, \gamma^\mu \, \partial_\mu \, \nu_R{}_I
  -
  \Bigl(
  F_{\alpha I} \, \overline{L}_\alpha \, \Phi \, \nu_R{}_I
  + \frac{M_I}{2} \, \overline{\nu_R{}_I^c} \, \nu_R{}_I 
  + h.c.
  \Bigr)
\,,
\end{eqnarray}
where ${\cal L}_{\rm MSM}$ is the MSM Lagrangian.  $F_{\alpha I}$ are
neutrino Yukawa couplings, and $\Phi$ and $L_\alpha$ ($\alpha = e,
\mu, \tau$) are Higgs and lepton weak-doublets, respectively.  The
Majorana masses of right-handed neutrinos are denoted by $M_I$ which
are taken to be real and positive without loss of generality.  Here
and hereafter we work in a basis in which the mass matrix of charged
leptons is diagonal.  In this model, neutrinos also obtain the Dirac
masses, $[M_D]_{\alpha I} = F_{\alpha I} \vev{\Phi}$ ($\vev{\Phi}$ is
a vacuum expectation value of the Higgs field), after the electroweak
(EW) symmetry breaking.

The distinctive feature of the model is the region of the parameter
space of Eq.~(\ref{eq:L_nuMSM}), \ie, we restrict ourselves
in the region 
\begin{eqnarray}
  \left|[M_D]_{\alpha I} \right| \ll M_I \lesssim 100 \,\mbox{GeV} \,.
\end{eqnarray}
In this case the seesaw mechanism works, and mass eigenstates of
neutrinos are then divided into two groups.  One group consists of
active neutrinos $\nu_i$ ($i= 1,2,3$).  Their masses are found from
the seesaw mass matrix $M_\nu = - M_D M_I^{-1} M_D^T$ as
\begin{eqnarray}
  U^\dagger \, M_\nu \, U^\ast = \mbox{diag}(m_1, m_2, m_3) \,,
\end{eqnarray}
where $U$ is the mixing matrix of active neutrinos.  The other one
consists of sterile neutrinos $N_I$ which are almost the
right-handed states $N_I \simeq \nu_R{}_I$, and their masses are
approximately given by the Majorana masses $M_I$.  We then find the
neutrino mixing as
\begin{eqnarray}
  \nu_L{}_\alpha = 
  U_{\alpha i} \, \nu_i + \Theta_{\alpha I} \, N_I^c \,,
\end{eqnarray}
where $\Theta_{\alpha I} = [M_D]_{\alpha I}/M_I$ are the
active-sterile (left-right) mixing matrix.  We should stress that
sterile neutrinos $N_I$ here are originated from the right-handed
neutrinos in the seesaw mechanism.  Thus, we simply say
$N_I$ as right-handed neutrinos from now on.

In the $\nu$MSM three right-handed neutrinos play important roles in
cosmology.  One of them, say $N_1$, is a candidate for dark matter of
the universe~\cite{Asaka:2005an}.  It is beyond the scope of the
present work to describe this issue.  However, one thing being
important for later discussions is that the Yukawa couplings of $N_1$
should be highly suppressed to realise a successful dark matter
scenario. (See the details, \eg,
Refs.~\cite{Asaka:2006nq,Shaposhnikov:2008pf,Laine:2008pg,Boyarsky:2009ix}.)
As a result, the contribution from $N_1$ to the seesaw matrix $M_\nu$
becomes negligible~\cite{Asaka:2005an}.  Furthermore, as shown in
Ref.~\cite{Asaka:2005pn}, $N_1$ plays essentially no role to generate
BAU.  Therefore, we take $F_{\alpha 1} = 0$ for simplicity in the rest
of this analysis.

The other right-handed neutrinos, $N_2$ and $N_3$, are then
responsible to the masses and mixing angles of active neutrinos.
Notice that the lightest active neutrino becomes massless in our
approximation.  Further, the flavour oscillation between $N_2$ and
$N_3$ in the early universe can be a source of BAU through the
mechanism proposed in Ref.~\cite{Akhmedov:1998qx}, as we will show below.
In the $\nu$MSM, therefore, BAU is related to the parameters of
active neutrinos through physics of $N_2$ and $N_3$.

The neutrino Yukawa matrix $F$ for $N_2$ and $N_3$, which is a $3
\times 2$ matrix, can be expressed without loss of generality
as~\cite{Casas:2001sr}
\begin{eqnarray}
  \label{eq:F}
    F = \frac{i}{\vev{\Phi}} \,
    U \, D_\nu^{1/2} \, \Omega \, D_N^{1/2} \,.
\end{eqnarray}
Here parameters of active neutrinos are their 
masses $D_\nu = \mbox{diag}(m_1, m_2,m_3)$ and 
the mixing matrix
\begin{eqnarray}
  U = 
  \left( 
    \begin{array}{c c c}
      c_{12} c_{13} &
      s_{12} c_{13} &
      s_{13} e^{- i \delta} 
      \\
      - c_{23} s_{12} - s_{23} c_{12} s_{13} e^{i \delta} &
      c_{23} c_{12} - s_{23} s_{12} s_{13} e^{i \delta} &
      s_{23} c_{13} 
      \\
      s_{23} s_{12} - c_{23} c_{12} s_{13} e^{i \delta} &
      - s_{23} c_{12} - c_{23} s_{12} s_{13} e^{i \delta} &
      c_{23} c_{13}
    \end{array}
  \right)  
  \times
  \mbox{diag} 
  ( 1 \,,~ e^{i \eta} \,,~ 1) \,,
\end{eqnarray}
with $s_{ij} = \sin \theta_{ij}$ and $c_{ij} = \cos \theta_{ij}$.  The
Dirac and Majorana phases are denoted by $\delta$ and $\eta$,
respectively.  Since we set $F_{\alpha 1} = 0$, the masses of active
neutrinos are
\begin{eqnarray}
  &&m_3 = m_{\rm atm} > m_2 = m_{\rm sol} > m_1 =0
  ~~~\mbox{in the NH case} \,,
  \nonumber \\
  &&m_2 = \sqrt{m_{\rm atm}^2 + m_{\rm sol}^2} > 
  m_1 = \sqrt{m_{\rm atm}^2} > m_3 = 0
  ~~~\mbox{in the IH case} \,,
\end{eqnarray}
The observational data of mixing angles are $s_{12}^2 =
0.318^{+0.062}_{-0.048}$, $s_{23}^2 = 0.50^{+0.17}_{-0.14}$, and
$s_{13}^2 \le 0.053$, respectively,
and masses are 
$m_{\rm sol}^2 = \Delta m_{21}^2 = (7.59^{+0.68}_{-0.56}) \times 10^{-5}
\eV^2$, and 
$m_{\rm atm}^2 = |\Delta m_{31}^2| = (2.40^{+0.35}_{-0.33} )\times
10^{-3}\eV^2$ (at the $3 \sigma$ level)~\cite{Schwetz:2008er}.
Hereafter, we shall adopt the central values
unless otherwise stated.

On the other hand, parameters of $N_2$ and $N_3$ are
their masses $D_N = \mbox{diag}(M_2,M_3)$ and the $3 \times 2$
matrix
\begin{eqnarray}
  &&\Omega =
  \left(
    \begin{array}{c c}
      0 & 0 \\
      \cos \omega & - \sin \omega \\
      \xi \sin \omega & \xi \cos \omega
    \end{array}
  \right)
  ~~\mbox{in the NH case} \,,
  \nonumber \\
  &&\Omega =
  \left(
    \begin{array}{c c}
      \cos \omega & - \sin \omega \\
      \xi \sin \omega & \xi \cos \omega \\
      0 & 0 
    \end{array}
  \right)
  ~~\mbox{in the IH case} \,,
\end{eqnarray}
where $\xi = \pm 1$ and $\omega$ is an arbitrary complex number.

In the considering situation there are three CP violating parameters,
$\delta$, $\eta$ and $\mbox{Im}\omega$, in leptonic sector, which can
potentially contribute to the generation of BAU.  
In this analysis we concentrate
on the case in which BAU is originated solely from the CP phases in
the mixing matrix of active neutrinos, and find the dependence on
$\delta$ and $\eta$ as well as the mixing angles $\theta_{ij}$ by
taking $\mbox{Im}\omega =0$.%
\footnote{
The impact of $\mbox{Im}\omega$ on BAU will be discussed 
in elsewhere~\cite{AsakaIshida}.}

\section{Baryogenesis via Neutrino Oscillation}
Let us then discuss how BAU is generated in the $\nu$MSM through
baryogenesis via oscillation of right-handed
neutrinos~\cite{Akhmedov:1998qx}.  In the considering model the
lepton-number violations due to Majorana masses are ineffective for
high temperatures $T \gtrsim 100$ GeV~\cite{Akhmedov:1998qx}.  In
order to generate the baryon asymmetry, thus, it is crucial the lepton
asymmetry is distributed into left-handed leptons $L_\alpha$ and
right-handed neutrinos $N_I$ for the temperatures of interest rather
than its creation.  Then, the asymmetry stored in the left-handed
sector is partially transferred into the baryon asymmetry due to the
$B+L$ breaking sphaleron transition which is rapid for $T > T_W \simeq
100$ GeV~\cite{Kuzmin:1985mm}.

We denote the asymmetries of numbers of $N_{2,3}$ and $L_\alpha$
by $\Delta N_{2,3}$ and $\Delta L_\alpha$, respectively.
These asymmetries are estimated by solving the
kinetic equations for their density matrices $\rho_{NN}$ and
$\rho_{LL}$%
\footnote{
The density matrix $\rho_{LL}$ denotes the 
sum of $\rho_{\nu_L \nu_L}$ and $\rho_{e_L e_L}$,
which are the same in the temperatures under consideration
due to SU(2)$_L$ symmetry.
},
 which are given by
\begin{eqnarray}
  \label{eq:EOM_N}
  i\frac{d \rho_{NN}}{dt} 
  &=& [H_{NN}^0 + V_N, \rho_{NN}] -
  \frac{i}{2} \{ \Gamma_{N}, \rho_{NN} - \rho_{NN}^{eq} \}
  + \frac{i \sin \phi}{8} T 
  F^\dagger (\rho_{LL} - \rho_{LL}^{eq} ) F \,,
  \\
  \label{eq:EOM_L}
  i\frac{d \rho_{LL}}{dt} 
  &=& [H_{LL}^0 + V_L, \rho_{LL}] -
  \frac{i}{2} \{ \Gamma_{L}, \rho_{LL} - \rho_{LL}^{eq} \}
  + \frac{i \sin \phi}{4} T 
  F (\rho_{NN} - \rho_{NN}^{eq} ) F^\dagger \,,
\end{eqnarray}
where note again that $F$ is the $3 \times 2$ matrix for neutrino
Yukawa couplings of $N_2$ and $N_3$.
$H_{NN}^0$ and $H_{LL}^0$ denote the Hamiltonian when 
$F=0$.  The effective potentials 
and the destruction rates for $N$ and $L_\alpha$ are 
\begin{eqnarray}
  \label{eq:GAM_N}
  V_N \eqn{=} \frac{1}{8} \, T \, F^\dagger F \,,~~~~  
  \Gamma_N = 2 \, s_\phi \, V_N \,,
  \\
  V_L \eqn{=} \frac{1}{16} \, T \, F^\dagger F \,,~~~~
  \Gamma_L = 2 \, s_\phi \, V_L \,.
\end{eqnarray}
where $s_\phi \simeq 2 \times 10^{-2}$~\cite{Akhmedov:1998qx}.
Notice that these expressions are valid as long as 
$T$ is sufficiently higher than $T_W$.
The kinetic equations of the density matrices 
for the anti-particles
$\rho_{\bar N \bar N}$ and $\rho_{\bar L \bar L}$
are obtained by replacing $F \to F^\ast$ in Eqs.~(\ref{eq:EOM_N})
and (\ref{eq:EOM_L}).

These equations include the medium effects of surrounding hot plasma,
{\it i.e.}, the thermal potentials which describe the coherent
oscillations of right-handed neutrinos induced by $V_N$,
 and the decoherent terms which
describe the production and destruction of
$N_I$~\cite{Akhmedov:1998qx}.  Furthermore, we include the terms which
express the exchange of asymmetries between left and right-handed
sectors~\cite{Asaka:2005pn}.  
It is found from Eqs.~(\ref{eq:EOM_N}) and (\ref{eq:EOM_L}) that
$\Delta N_{\rm tot} + \Delta L_{\rm tot} =0$, 
where $\Delta N_{\rm tot} = \sum_{I=2,3} \Delta N_I$
and $\Delta L_{\rm tot} = \sum_{\alpha = e, \mu, \tau} \Delta L_\alpha$,
which is crucial in the considering baryogenesis scenario
as mentioned above.
For further details of these issues,
please see Ref.~\cite{Asaka:2005pn}.

The coupled equations (\ref{eq:EOM_N}) and (\ref{eq:EOM_L}) can be
solved not only numerically, but also analytically by using the
perturbative expansion of the Yukawa coupling constants
$F$~\cite{Asaka:2005pn}.  
The initial conditions are taken as 
$\rho_{NN}(0)=\rho_{\bar N \bar N}(0)=0$ and
$\rho_{LL}(0)=\rho_{\bar L \bar L}(0)=\rho_{LL}^{eq}$.
Then, we can estimate the asymmetries as
$\Delta N_I = [\rho_{NN}]_{II} - [\rho_{\bar N \bar N}]_{II}$ and
$\Delta L_\alpha = [\rho_{LL}]_{\alpha \alpha} - [\rho_{\bar L \bar
  L}]_{\alpha \alpha}$.  In the following we will present the 
analytical expression of the active flavour asymmetry 
as well as BAU at the leading order of $F$.  We will
also show the results from the numerical solutions of
Eqs.~(\ref{eq:EOM_N}) and (\ref{eq:EOM_L}), which confirm the validity
of the analytical expressions of the asymmetries.

\section{Active Flavour Asymmetries}
First of all, we discuss the yield 
of the active flavour asymmetry $\Delta L_\alpha$ 
($\alpha = e, \mu, \tau)$.
The leading order contribution to $\Delta L_\alpha$
is induced at ${\cal O}(F^4)$ and 
the analytic expression 
at the temperature $T$
can be written as~\cite{Asaka:2005pn}
\begin{eqnarray}
  \Delta L_\alpha (T) 
  = \frac{ s_\phi^2}{4} \,
  A_{32}^\alpha \, 
  \frac{M_0^2}{T_L^2} \,
  J_{32} (T_L/T) \,,
\end{eqnarray}
where $M_0 \simeq 7.1 \times 10^{17}$ GeV and 
the CP asymmetry parameter $A_{32}^\alpha$ is defined by
\begin{eqnarray}
  A_{32}^\alpha 
  \eqn{=}
  \mbox{Im}
  \Bigl[ 
  F_{\alpha 3} \, [F^\dagger F]_{32} \, F_{\alpha 2}^\ast 
  \Bigr] \,.
\end{eqnarray}
The typical temperature of the generation of
$\Delta L_\alpha$, $T_L$, is given by
\begin{eqnarray}
  \label{eq:T_L}
  T_L 
  = \left ( \frac{1}{6} \, M_0 \, \Delta M_{32}^2 \right)^{\frac 13} 
  =
  2.2 \, \TeV
  \left( \frac{M_3}{3 \GeV} \right)^{\frac 23}
  \left( \frac{\Delta M_{32}^2/M_3^2}{10^{-8} } \right)^{\frac 13}
  \,,
\end{eqnarray}
where $\Delta M_{32}^2 = M_3^2 - M_2^2$.
The evolution of $\Delta L_\alpha$ is described by 
the function $J_{32} $
\begin{eqnarray}
  J_{32} (x) = 
  \int_0^{x} dx_1 \int_0^{x_1} dx_2 
  \sin (x_1^3 - x_2^3 ) \,.
\end{eqnarray}
It is found that $J_{32} (x) \simeq \frac{3}{20} x^5$ 
for $x \ll 1$ while it is oscillating around and
approaching to the value
\begin{eqnarray}
  J_{32}(x)|_{x \gg 1}
  =  \frac{2^{1/3} \pi^{3/2}}{9 \Gamma (5/6)}  
  =  0.69 \,.
\end{eqnarray}
Therefore, the active flavour asymmetry scales as
$\Delta L_\alpha \propto T^5$ for $T \gg T_L$ and
takes a constant value for $T \ll T_L$ as
\begin{eqnarray}
  \Delta L_\alpha 
  \eqn{=}
  \frac{3^{2/3} \, \pi^{3/2} \, s_\phi^2}{18 \, \Gamma(5/6)} \,
  A_{32}^\alpha \,
  \frac{ M_0^{4/3} }{(\Delta M_{32}^2)^{2/3}} \,.
\end{eqnarray}
We can see that $\Delta L_\alpha$ for $T \ll T_L$
is enhanced when $N_2$ and $N_3$ are quasi 
degenerate~\cite{Asaka:2005pn}.

Now we would like to express the CP asymmetry parameter
$A_{32}^\alpha$ by using the parameters of the neutrino Yukawa couplings in
Eq.~(\ref{eq:F}).  In the NH case we evaluate $A^\alpha_{32}$ as
\begin{eqnarray}
  \label{eq:DELAL_NH}
  A_{32}^\alpha
  \eqn{=}
  \xi \, \sin 2 \mbox{Re}\omega \,
  \frac{M_3 M_2 (m_3 m_2)^{1/2} (m_3 - m_2)}{4 \, \vev{\Phi}^4}
  \times a_\alpha
  \nonumber \\
  \eqn{=}
  \xi \, \sin 2 \mbox{Re}\omega \,
  \frac{M_3 M_2 m_{\rm atm}^{3/2} m_{\rm sol}^{1/2} }{4 \, \vev{\Phi}^4}
  \times S_{m_\nu}
  \times a_\alpha 
  \,,
\end{eqnarray}
where $S_{m_\nu}$ is
\begin{eqnarray}
  \label{eq:Smnu_NH}
  S_{m_\nu} = 1 - r_m \,,
\end{eqnarray}
where $r_m = m_{\rm sol}/m_{\rm atm}$.
The parameter $a_\alpha$ can be evaluated as
\begin{eqnarray}
  \label{eq:A_NH}
  a_e \eqn{=}
  - \sin \theta_{12} \sin 2 \theta_{13} \sin ( \delta + \eta) \,,
  \nonumber \\
  a_\mu \eqn{=}
  + \sin^2 \theta_{23} \sin \theta_{12} \sin 2 \theta_{13}
  \sin ( \delta + \eta) 
  - \sin 2 \theta_{23} \cos \theta_{12} \cos \theta_{13}
  \sin \eta \,,
  \nonumber \\
  a_\tau \eqn{=}
  + \cos^2 \theta_{23} \sin \theta_{12} \sin 2 \theta_{13}
  \sin ( \delta + \eta) 
  + \sin 2 \theta_{23} \cos \theta_{12} \cos \theta_{13}
  \sin \eta \,.
\end{eqnarray}
We find that the active flavour asymmetry depends on parameters of
right-handed neutrinos as $\Delta L_\alpha \propto \xi 
\sin \mbox{2Re}\omega \, M_2 M_3/(\Delta M_{32}^2)^{2/3}$, 
and hence vanishes when $\mbox{Re}\omega = n \pi/2$ ($n$ is integer).
This is simply because the flavour oscillation between
$N_2$ and $N_3$, which is essential to the considering
mechanism, disappears.  On the other hand, the dependence on
the mixing angles and CP violating phases of active neutrinos is
summarised in $a_\alpha$.

As already pointed out in Ref.~\cite{Asaka:2005pn}, the total
asymmetry of active flavours vanishes $\Delta L_{\rm tot}
=0$ at the leading ${\cal O}(F^4)$ since $\sum_\alpha A_{32}^\alpha =
\mbox{Im} \Bigl[ [ F^\dagger F]_{23} \, [F^\dagger F]_{32} \Bigr] =
0$, which can be seen as $a_e + a_\mu + a_\tau = 0$ in
Eq.~(\ref{eq:A_NH}).  It is interesting to note that the active
flavour asymmetries depend on CP violating phases in two ways, {\it
  i.e.}, $\sin (\delta + \eta)$ and $\sin \eta$, and $\Delta L_e$ only
depends on the former one.  We also find that $\Delta L_e$ vanishes
and $\Delta L_\mu = - \Delta L_\tau$ when $\theta_{13} =0$.  In this
case $a_e =0$ and $a_\mu = - a_\tau = - \sin 2 \theta_{23} \cos
\theta_{12} \sin \eta$, and the asymmetries only depend on Majorana
phase $\eta$ as expected.

On the other hand, in the inverted hierarchy case, 
the CP asymmetry parameter $A^\alpha_{32}$ can be written as
\begin{eqnarray}
  A_{32}^\alpha
  \eqn{=}
  \xi \, \sin  2 \mbox{Re}\omega \,
  \frac{M_3 M_2 (m_2 m_1)^{1/2} (m_2 - m_1)}{4 \, \vev{\Phi}^4}
  \times a_\alpha
  \nonumber \\
  \eqn{=}
  \xi \, \sin  2 \mbox{Re}\omega \,
  \frac{M_3 M_2 m_{\rm atm}^{3/2} m_{\rm sol}^{1/2} }{4 \, \vev{\Phi}^4}
  \times S_{m_\nu} \times a_\alpha 
  \,,
\end{eqnarray}
where $a_\alpha$ is found as
\begin{eqnarray}
  a_e \eqn{=}
  + \sin 2 \theta_{12} \cos^2 \theta_{13} \sin \eta \,,
  \nonumber \\
  a_\mu \eqn{=}
  - \Bigl[
  \sin^2 \theta_{23} \cos^2 \theta_{13} + \cos 2 \theta_{23}
  \Bigr] \sin 2 \theta_{12} \sin \eta 
  \nonumber \\
  \eqn{}
  + \sin 2 \theta_{23} \sin \theta_{13} 
  \Bigl[
    \cos \eta \sin \delta - 
    \cos 2 \theta_{12} \sin \eta \cos \delta
  \Bigr] \,,
  \nonumber \\
  a_\tau \eqn{=}
  - \Bigl[
  \cos^2 \theta_{23} \cos^2 \theta_{13} - \cos 2 \theta_{23}
  \Bigr] \sin 2 \theta_{12} \sin \eta
  \nonumber \\
  \eqn{}
  - \sin 2 \theta_{23} \sin \theta_{13} 
  \Bigl[
    \cos \eta \sin \delta - 
    \cos 2 \theta_{12} \sin \eta \cos \delta
  \Bigr] \,.
\end{eqnarray}
Further, we have introduced
\begin{eqnarray}
  \label{eq:Smnu_IH}
  S_{m_\nu} =
  r_m^{- \frac 12} ( 1 + r_m^2 )^{\frac 14} 
  \bigl[ ( 1 + r_m^2 )^{\frac 12} - 1 \bigr]
  = \frac{r_m^{\frac 32}}{2} \left[ 1 - \frac{r_m^4}{32} 
    + {\cal O}(r_m^6 ) \right] \,.
\end{eqnarray}
We find that the active flavour asymmetry in the IH 
depends on the parameters of right-handed neutrinos in the same 
way as in the NH.  The total asymmetry of active flavours 
at ${\cal O}(F^4)$ vanishes $\Delta L_{\rm tot} =0$
as in the NH case.

It should be noted that, comparing with the NH case
(\ref{eq:DELAL_NH}), $A_{32}^\alpha$ in the IH case receives of
suppression factor of $S_{m_\nu} \simeq 0.04$, and thus the
production of active flavour asymmetries in IH is less effective than
NH apart from $a_\alpha$.  Further, $\Delta L_\alpha$
depends on CP violating phases differently from the NH case.  When $\theta_{13}
= 0$, we find that
\begin{eqnarray}
  a_e \eqn{=} + \sin 2 \theta_{12} \, \sin \eta \,,
  \nonumber \\
  a_\mu \eqn{=}
  - \sin 2 \theta_{12} \cos^2 \theta_{23} \, \sin \eta \,,
  \nonumber \\
  a_\tau \eqn{=}
  - \sin 2 \theta_{12} \sin^2 \theta_{23} \, \sin \eta \,.
\end{eqnarray}
Thus, as in the NH case, the asymmetries only depend
on Majorana phase $\eta$, as expected.  In this case, however, 
$\Delta L_e$ does not vanish even when $\theta_{13}=0$,
and we have $\Delta L_\mu = \Delta L_\tau = - 0.5 \Delta L_e$
for $\theta_{23} = \pi/4$.

The total asymmetry $\Delta L_{\rm tot}$ is 
induced at ${\cal O}(F^6)$ and $\Delta L_{\rm tot} = - 
\Delta N_{\rm tot}$.  
It is found in Ref.~\cite{Asaka:2005pn} that 
\begin{eqnarray}
  \Delta L_{\rm tot} (T)
  \eqn{=}
  {} - \frac{s_\phi}{8} \int_T^\infty dT_1 
  \frac{M_0}{T_1^2} \,
  \sum_{I=2,3} \sum_{\alpha = e, \mu,\tau} 
  \abs{F_{I\alpha}^2} \Delta L_\alpha (T)
  \nonumber \\
  \eqn{=}
  {} -
  \frac{s_\phi^3}{32} 
  \sum_{I=2,3} \sum_{\alpha = e, \mu,\tau} 
  \abs{F_{I\alpha}^2} \, A_{32}^\alpha \,
  \frac{M_0^3}{T_L^3} \, K(T_L/T) \,,
\end{eqnarray}
where the evolution of $\Delta L_{\rm tot}$ is 
described by the function $K$:
\begin{eqnarray}
  K (x ) = \int_0^x dx_1 \, J_{32}(x_1) \,.
\end{eqnarray}
We find that $K(x) \simeq x^6/40$ for $x \ll 1$
while $K(x) \simeq J_{32}(\infty) \, x$ for $x \gg 1$.%
\footnote{
We numerically find that $K(x)$ can be fitted as
$K(x) = a x - b$ for $2 \lesssim x \lesssim 100$,
where $a = 0.68 \simeq J_{32}(\infty)$ and $b=1.4$.
}
\begin{figure}[t]
  \centerline{
  \includegraphics[scale=1]{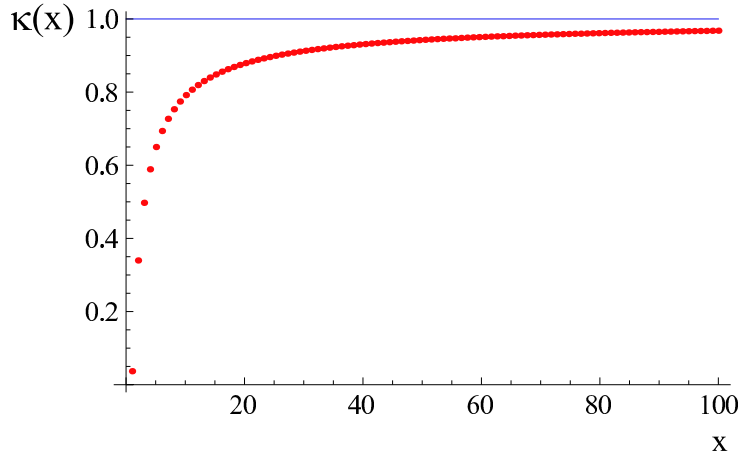}
  }%
  \caption{ \label{fig:FUNK_KAP}
    Function $\kappa (x)$ in Eq.~(\ref{eq:kappa}).
    }
\end{figure}
Then, we normalise $K(x)$ as
\begin{eqnarray}
  \label{eq:kappa}
  K(x) = J_{32}(\infty) \, x \, \kappa (x) \,,
\end{eqnarray}
where the behaviour of the function $\kappa (x)$ is 
shown in Fig.~\ref{fig:FUNK_KAP}.

It should be noted that we have so far assumed that $N_2$ and $N_3$
are out of equilibrium in order to avoid the wash-out of the
asymmetries.  When the interaction rate of right-handed neutrinos
(\ref{eq:GAM_N}) is $\Gamma_N < H$ (the Hubble parameter) till the
sphaleron transitions are switched off at $T = T_W \simeq 100\GeV$,
the eigenvalues of $F^\dagger F$ should be smaller than $2.8 \times
10^{-14}$.  This is translated into the upper bound on the mass of
right-handed neutrinos as $M_{2,3} \lesssim 17
\GeV$~\cite{Asaka:2005pn} for both the NH and IH cases by taking $N_2$
and $N_3$ are quasi-degenerate, which is required to explain the
observed BAU as we will show.  We thus restrict ourselves in this mass
region below.

\section{Baryon Asymmetry of the Universe}
Now we are at the position to present the analytical expression of BAU
and study how it depends on masses, mixing angles and CP violating
phases of active neutrinos.  The baryon-to-entropy ratio, $Y_B =
n_B/s$ ($n_B$ and $s$ are the baryon and entropy densities at the
present universe, respectively), is obtained from $\Delta L_{\rm tot}$
as
\begin{eqnarray}
  Y_B = - c_s \, \frac{28}{79} \, 
  \Delta L_{\rm tot} (T_W) \,,
\end{eqnarray}
where $c_s = 7.0 \times 10^{-4}$.
Notice that only the asymmetry in the left-handed leptons
is transferred via the sphaleron processes
into the baryon asymmetry as
$\Delta B = - \frac{28}{79} \Delta L_{\rm tot}$~\cite{Khlebnikov:1988sr},
and this transition is switched off for $T < T_W$.

We take the typical temperature $T_L$ (\ref{eq:T_L}) for the
generation of $\Delta L_\alpha$ as $T_L \gtrsim T_W$, and $N_2$ and
$N_3$ are quasi-degenerate, {\ie} $\Delta M \ll M_N$ where $M_3 = M_N
+ \Delta M/2$ and $M_2 = M_N - \Delta M/2$.  These are arranged in
order to enhance the production of the baryon asymmetry.  In this
case, we find that the analytical expression of $Y_B$ 
for both the NH and IH cases is given by
\begin{eqnarray}
  Y_B 
  \eqn{=}
  1.8 \times 10^{-11} \,
  \kappa (T_L/T_W) \,
  \frac{M_0^{7/3} M_N^{5/3} m_{\rm atm}^{5/2} m_{\rm sol}^{1/2}}
  {T_W (\Delta M^2_{32}/M_N^2)^{2/3} \vev{\Phi}^6}
  \times \delta_{\rm CP}
  \nonumber \\
  \eqn{=}
  4.7 \times 10^{-10} \, \delta_{\rm CP}
  \left(
    \frac{10^2 \GeV}{T_W}
  \right)
  \left(
    \frac{M_N}{5\GeV}
  \right)^{5/3}
  \left(
    \frac{10^{-8}}{\Delta M_{32}^2/M_N^2}
  \right)^{2/3} \,,
\end{eqnarray}
where we have taken $\kappa (T_L/T_W) = 1$ in the 
last equality.  The CP asymmetry parameter $\delta_{\rm CP}$ for BAU 
is expressed as
\begin{eqnarray}
  \label{eq:DEL_CP}
  \delta_{\rm CP} 
  \eqn{=} \xi \, \sin 2 \mbox{Re} \omega \cdot
  S_{m_\nu} \,  \delta_\nu \,.
\end{eqnarray}
Here $S_{m_\nu}$ is the parameter concerning with the neutrino 
mass hierarchy in Eq.~(\ref{eq:Smnu_NH}) or (\ref{eq:Smnu_IH})
for the NH or IH case, respectively.
Therefore, $Y_B$ in the IH case is suppressed by 
about 4\% compared with the NH case apart from $\delta_\nu$.

It can be seen that to get the sizable $Y_B$ the mass degeneracy of
$N_2$ and $N_3$ at a rather high accuracy is required.  Note, however,
that such a small mass difference is stable against the radiative
corrections due to the smallness of Yukawa coupling constants of
neutrinos under consideration.  We also find that
a larger $Y_B$ can be obtained for a larger $M_N$ 
as long as $M_N \lesssim 17$ GeV.  Otherwise, $N_2$ and 
$N_3$ get in thermal equilibrium and the asymmetries are
washed out.  It should be noted that
$Y_B$ is proportional to $\xi\sin \mbox{Re}\omega$, which means that
the sign of $Y_B$ cannot be uniquely predicted
even when all the parameters of active neutrinos are experimentally
determined.

In this expression we have introduced 
the CP asymmetry parameter $\delta_\nu$ to 
describe how $Y_B$ depends on the mixing angles 
and CP violating phases in the mixing matrix $U$.
The analytical expression for $\delta_\nu$ for both the NH and IH 
cases are presented in Appendix~\ref{sec:AP}.
From now on we shall study these dependence by using 
the analytical expression as well as the numerical estimation
of $Y_B$, which is obtained by solving numerically 
Eqs.~(\ref{eq:EOM_N}) and (\ref{eq:EOM_L}).

\begin{figure}[t]
  \includegraphics[scale=1.3]{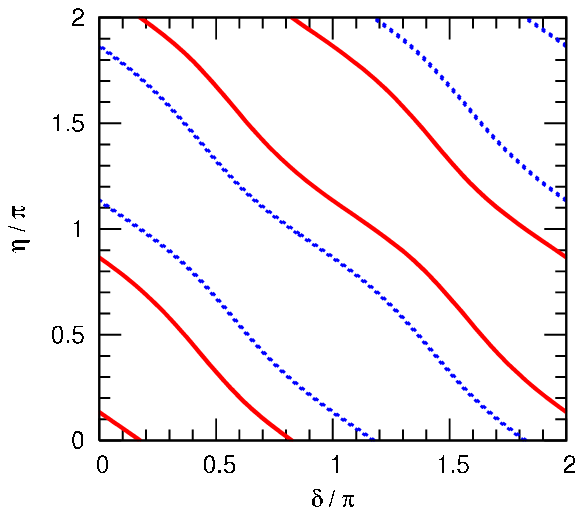}%
  \includegraphics[scale=1.3]{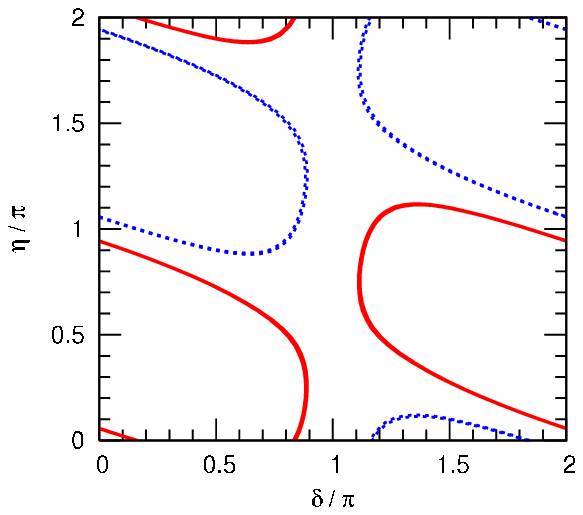}%
  \caption{ \label{fig:FIG_DELETA_NH_TH23}
    Parameter regions in the $\delta$-$\eta$ plane
    accounting for the observed baryon asymmetry
    in the NH case are shown by lines.
    The red solid lines are for $\xi = +1$ while the blue
    dashed lines are for $\xi = -1$.
    Here we take $M_3 = 5$ GeV,
    $\Delta M_{32}^2/M_3^2 = 10^{-8}$, $\mbox{Re}\omega =\pi/4$,
    and 
    $\sin^2 \theta_{13} = 0.053$.
    We take $\sin^2 \theta_{23} = 0.5$ (left) and
    $0.36$ (right), respectively.
    }
\end{figure}
In the NH case we find that the leading term 
of $\delta_\nu$ in the expansion of $r_m$ is given by
\begin{eqnarray}
  \label{eq:DEL_NH}
  \delta_{\nu} 
  \eqn{=}
  \frac{1}{2} 
  \sin \theta_{12} \sin 2 \theta_{13}
  [ \cos^2 \theta_{13} (3+\cos 4\theta_{23})- 4 \sin^2 \theta_{13} ]
  \,
  \sin (\delta + \eta)
  \nonumber \\
  \eqn{}
  +
  \cos \theta_{12} \sin 4 \theta_{23} \cos^3 \theta_{13} \sin \eta
  + {\cal O}(r_{\rm m}) 
  \,.
\end{eqnarray}
It is seen that $\delta_\nu$ depends on the CP violating phases
in two ways, \ie, $\delta + \eta$ and $\eta$.

We then find that when the mixing angle in the atmospheric neutrino
oscillation is maximal $\theta_{23} = \pi/4$, the leading term of
$\delta_\nu$ depends only on the sum of the CP phases $\delta + \eta$.
This behaviour can be understood in the left panel of
Fig.~\ref{fig:FIG_DELETA_NH_TH23}, which represents the parameter
region accounting for the present observation data of $Y_B$
in the $\delta$-$\eta$ plane obtained by 
the numerical solutions of Eqs.~(\ref{eq:EOM_N}) and 
(\ref{eq:EOM_L}).
When $\theta_{23}$ is slightly smaller than $\pi/4$,
the allowed region becomes drastically changed as
shown in the right panel of Fig.~\ref{fig:FIG_DELETA_NH_TH23}.
Thus, the deviation of $\theta_{23}$ from $\pi/4$, 
which will be tested in future oscillation experiments, 
is significant to determine $Y_B$ in the considering case.

On the other hand, when $\theta_{13} = 0$, 
the CP asymmetry parameter becomes
\begin{eqnarray}
  \delta_\nu = \sin 4 \theta_{23} \cos \theta_{12}
  (1 - r_m \cos^2 \theta_{12} ) \sin \eta \,,
\end{eqnarray}
including the higher order term of $r_m$.
In this case the asymmetry depends only on the Majorana phase 
$\eta$ as expected (since the Dirac phase $\delta$ always
appears together with $s_{13}$).  It is very important to 
note that $\delta_\nu =0$ when $\theta_{13} =0$ and $\theta_{23} = \pi/4$.
In this case, the generation of BAU in the NH case 
is ineffective and $Y_B$ at ${\cal O}(F^6)$ vanishes.

\begin{figure}[t]
  \includegraphics[scale=1.3]{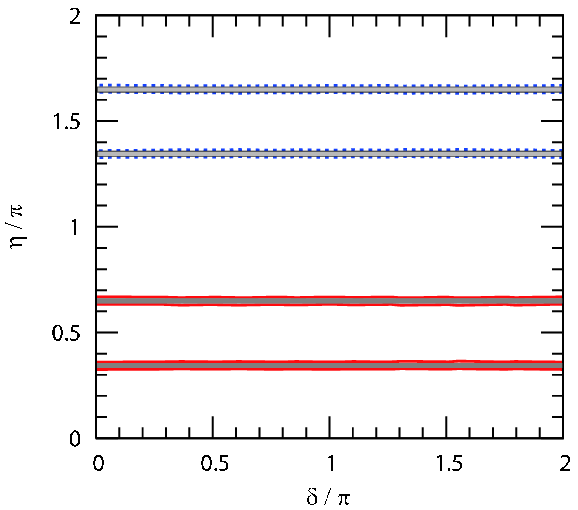}%
  \includegraphics[scale=1.3]{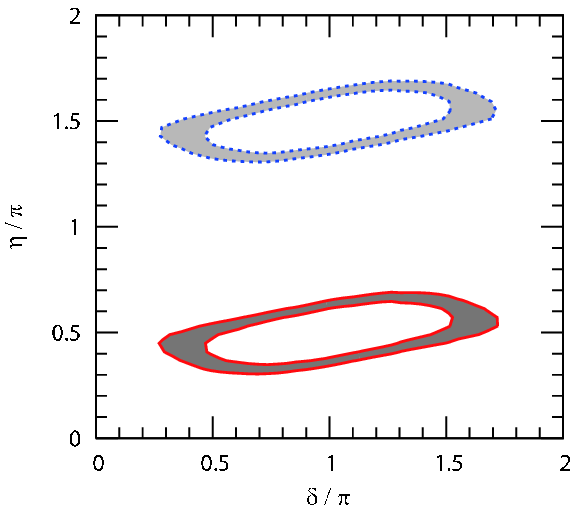}%
  \caption{ \label{fig:FIG_DELETA_IH_TH23}
    Parameter regions in the $\delta$-$\eta$ plane
    accounting for the observed baryon asymmetry
    in the IH case are shown by shaded regions.
    The red solid lines are for $\xi = +1$ while the blue
    dashed lines are for $\xi = -1$.
    Here we take $M_3 = 15$ GeV,
    $\Delta M_{32}^2/M_3^2 = 10^{-8}$, $\mbox{Re}\omega =\pi/4$,
    and 
    $\sin^2 \theta_{13} = 0.053$.
    We take $\sin^2 \theta_{23} = 0.5$ (left) and
    $0.36$ (right), respectively.
    }
\end{figure}
Next, we turn to consider the IH case, where the CP asymmetry
parameter $\delta_{\nu}$ at ${\cal O}(r_m^0)$ is estimated as
\begin{eqnarray}
  \label{eq:DEL_IH}
  \delta_{\nu} 
  \eqn{=}
  \frac{1}{4} \sin 2 \theta_{12}
  \cos^2 \theta_{13}
  \bigr[ 
  -5 -3 \cos 4 \theta_{23} + \cos 2 \theta_{13} (7+\cos 4 \theta_{23})
  \bigl]
  \sin \eta
  \nonumber \\
  \eqn{}
  +
  \sin 4 \theta_{23} \cos^2 \theta_{13} \sin \theta_{13} 
  (\sin \delta \cos \eta - \cos 2 \theta_{12} \cos \delta
  \sin \eta) + {\cal O}(r_m^2) 
  \,.
\end{eqnarray}
It is then found that $\delta_\nu$ at the leading order depends only
on the Majorana phase when $\theta_{23}=\pi/4$, which should be 
compared with the NH case.  This behaviour is shown in the left panel
of Fig.~\ref{fig:FIG_DELETA_IH_TH23}.
Moreover, we discover that $\delta_\nu$ in the IH case
does not vanish even when 
$\theta_{23} = \pi/4$ and $\theta_{13} =0$;
\begin{eqnarray}
  \delta_{\nu} 
  \eqn{=}
  \frac{1}{2}
  \left[
    1 + (1+ r_m^2)^{1/2} 
    + 3 [1-(1+ r_m^2)^{1/2} ]\cos 2\theta_{12}
  \right] \sin 2 \theta_{12} \sin \eta 
  \nonumber \\
  \eqn{=}
  \sin 2 \theta_{12} \sin \eta 
  \left[
    1 + \frac 14 ( 1 -3 \cos 2 \theta_{12} ) r_m^2
    + {\cal O}(r_m^4)
  \right] \,.
\end{eqnarray}
This is one important feature of generating BAU in the IH case.

\begin{figure}[t]
  \includegraphics[scale=1.3]{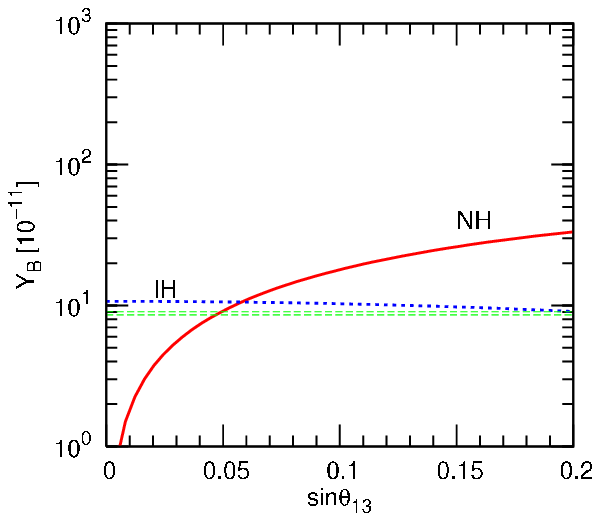}%
  \includegraphics[scale=1.3]{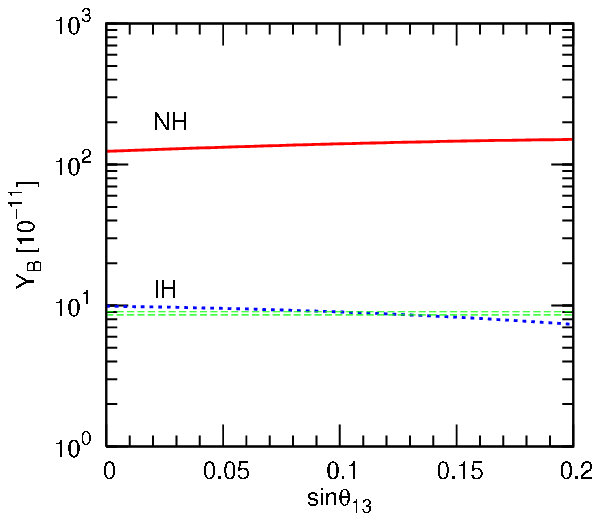}%
  \caption{ \label{fig:FIG_S13xpTH23}
    $Y_B$ in terms of $\sin \theta_{13}$ are shown by lines.
    The red solid line is for the NH case while
    the blue dashed line is for the IH case.
    The horizontal, green long dashed lines
    show the region for the observed baryon asymmetry.
    Here we take $M_3 = 15$ GeV,
    $\Delta M_{32}^2/M_3^2 = 10^{-8}$, $\mbox{Re}\omega =\pi/4$,
    $\delta = 7 \pi/4$, and $\eta = \pi/3$.
    We take $\sin^2 \theta_{23} = 0.5$ (left) and
    $0.36$ (right), respectively.
    }
\end{figure}
Therefore, $Y_B$ in the IH case is smaller than the NH case
in the most of the parameter space,
since it receives extra suppression factor of $S_{m_\nu}$ 
in Eq.~(\ref{eq:Smnu_IH}).  See Fig.~\ref{fig:FIG_S13xpTH23}.
However, when $\theta_{13}$ is very small and also $\theta_{23}$ 
is almost maximal, baryogenesis in the NH case
becomes ineffective and hence the IH of active neutrino 
masses is the essence of accounting for the present BAU
in the $\nu$MSM.  

\section{Conclusions}
\label{sec:Conc}
We have studied baryogenesis in the $\nu$MSM via flavour oscillation
between right-handed neutrinos $N_2$ and $N_3$.  In particular, the
case when BAU comes solely from the CP violating phases in the mixing
matrix of active neutrinos has been investigated.  We have presented
the analytical expressions of BAU for both the NH and IH cases of
active neutrino masses, and have demonstrated how the present value of
BAU depends on the Dirac and Majorana phases as well as the neutrino
mixing angles.  We have shown that BAU depends on the neutrino 
mass hierarchy and $Y_B$ in the IH case 
receives the suppression factor of $S_{m_\nu} \simeq 4\%$.
It has been found that $Y_B$ is very sensitive to 
the mixing angles $\theta_{23}$ and $\theta_{13}$.
When $\theta_{23} = \pi/4$, the leading contribution to $Y_B$
is proportional to $\sin (\delta + \eta)$ for the NH case
while to $\sin \eta$ for the IH case.
Moreover, when $\theta_{23}$ is almost maximal and 
$\theta_{13}$ is very small, the CP asymmetry parameter in 
$Y_B$ vanishes and no baryon asymmetry is generated 
(at least the leading ${\cal O}(F^6)$ contribution) in the NH case.
In this case, the IH case is required to explain 
the observed BAU.

\section*{Acknowledgments}
The work of T.A. was partially supported by the Ministry of Education,
Science, Sports and Culture, Grant-in-Aid for Scientific Research,
No.~21540260, and by Niigata University Grant for Proportion of
Project.  We thank Mikhail Shaposhnikov for correspondence
and reading of the manuscript.

\appendix
\section{CP Asymmetry Parameter $\delta_\nu$}
\label{sec:AP}
In this appendix, we write down the CP asymmetry parameter
$\delta_\nu$ defined in Eq.~(\ref{eq:DEL_CP}).
This parameter can be found from
\begin{eqnarray}
  \sum_{I=2,3} 
  \sum_{\alpha=e,\mu,\tau} |F_{\alpha I}|^2 \, a_\alpha
  \eqn{=}
  \frac{m_{\rm atm} \, M_N}{2 \,\vev{\Phi}^2} 
  \delta_\nu \,,
\end{eqnarray}
where we have neglected the terms which are proportional 
to $\Delta M/M_N$.

In the NH case, we can parameterize $\delta_\nu$ as
\begin{eqnarray}
  \delta_\nu = d_1 + r_m d_2 \,.
\end{eqnarray}
The leading term $d_1$ is estimated as
\begin{eqnarray}
  d_1 \eqn{=}
    \frac{1}{2} 
  \sin \theta_{12} \sin 2 \theta_{13}
  [ \cos^2 \theta_{13} (3+\cos 4\theta_{23})- 4 \sin^2 \theta_{13} ]
  \,
  \sin (\delta + \eta)
  \nonumber \\
  \eqn{}
  +
  \cos \theta_{12} \sin 4 \theta_{23} \cos^3 \theta_{13} \sin \eta
  + {\cal O}(r_{\rm m}) 
  \,,
\end{eqnarray}
which had already shown in Eq.~(\ref{eq:DEL_NH}).  For the sub-leading
term we can find that
\begin{eqnarray}  
  d_2 
  \eqn{=} 
  - 2 \sin \theta_{12} \cos^2 \theta_{12} \sin 4 \theta_{23} \cos \theta_{13} \left( 1 + \sin^2 \theta_{12} \cos 2 \theta_{13} \right) \sin \eta \nonumber\\
  \eqn{}- 
  \sin \theta_{12} \cos^2 \theta_{12} \sin^2 \theta_{23} \sin 2 \theta_{13} \sin \left( \delta - \eta \right) \nonumber\\
  \eqn{}+
  \frac{1}{8} \sin 2 \theta_{13} \left[-2 \sin^3 \theta_{12} \cos 2 \theta_{13} \left( 7 + \cos 4 \theta_{23} \right) 
  + \sin^2 2 \theta_{23} \left( \sin \theta_{12} + 5 \sin 3 \theta_{12} \right) \right] \sin \left( \delta + \eta \right) \nonumber\\
  \eqn{}+
  2 \sin^3 \theta_{12} \cos^2 \theta_{12} \sin 4 \theta_{34} \sin^2 \theta_{13} \cos \theta_{13} \sin \left( 2 \delta + \eta \right)\,.
\end{eqnarray}

In the IH case, on the other hand, we can parameterize 
$\delta_\nu$ as
\begin{eqnarray}
  \label{eq:DEL_IHAP}
  \delta_\nu = (1+ r_m^2)^{\frac 12} \, d_1 + d_2 \,,
\end{eqnarray}
where
\begin{eqnarray}
  d_1 
  \eqn{=}
  \Big[ \frac{1}{8} \sin 2 \theta_{12} \cos^2 \theta_{13} \left\{ -5 -3 \cos 4 \theta_{23} + \cos 2 \theta_{13} \left( 7 + \cos 4 \theta_{23} \right) \right\} \nonumber\\
  \eqn{}- \frac{1}{64}
  \sin 4 \theta_{12} \left\{ 29 + 27 \cos 4 \theta_{23} + \cos 4 \theta_{13} \left( 7 + \cos 4 \theta_{23} \right) + 40 \sin^2 \theta_{23} \cos 2 \theta_{13} \right\} \Big] \sin \eta \nonumber\\
  \eqn{}+ 
  \frac{1}{8} \sin 4 \theta_{23} \Big[ 4 \sin \theta_{13} \left\{ \cos^2 \theta_{13} + \cos \theta_{12} \left( 1 + \sin^2 \theta_{13} \right) \right\} \sin \delta \cos \eta \nonumber\\
  \eqn{}-
  \left\{ 4 \cos 2 \theta_{12} \sin \theta_{13} \cos^2 \theta_{13} + \cos 4 \theta_{12} \left( 7 \sin \theta_{13} - \sin 3 \theta_{13} \right) \right\} \cos \delta \sin \eta \Big] \nonumber\\
  \eqn{}-
  \frac{1}{2} \sin^2 2 \theta_{23} \sin^2 \theta_{13} \left( 2 \sin 2 \theta_{12} \sin 2 \delta \cos \eta - \sin 4 \theta_{12} \cos 2 \delta \sin \eta \right)\,,\\
  d_2
    \eqn{=}
  \Big[ \frac{1}{8} \sin 2 \theta_{12} \cos^2 \theta_{13} \left\{ -5 -3 \cos 4 \theta_{23} + \cos 2 \theta_{13} \left( 7 + \cos 4 \theta_{23} \right) \right\} \nonumber\\
  \eqn{}+ \frac{1}{64}
  \sin 4 \theta_{12} \left\{ 29 + 27 \cos 4 \theta_{23} + \cos 4 \theta_{13} \left( 7 + \cos 4 \theta_{23} \right) + 40 \sin^2 \theta_{23} \cos 2 \theta_{13} \right\} \Big] \sin \eta \nonumber\\
  \eqn{}+ 
  \frac{1}{8} \sin 4 \theta_{23} \Big[ \left\{ \sin \theta_{13} \left( 1 - 7 \cos 2 \theta_{12} \right) +2 \cos^2 \theta_{12} \sin 3 \theta_{13} \right\} \sin \delta \cos \eta \nonumber\\
  \eqn{}-
  \left\{ 4 \cos 2 \theta_{12} \sin \theta_{13} \cos^2 \theta_{13} + \cos 4 \theta_{12} \left( -7 \sin \theta_{13} + \sin 3 \theta_{13} \right) \right\} \cos \delta \sin \eta \Big] \nonumber\\
  \eqn{}+
  \frac{1}{4} \sin^2 \theta_{13} \left[ \sin 2 \theta_{12} \sin^2 2 \theta_{23} \sin 2 \delta \cos \eta - \sin 4 \theta_{23} \left( 1 - \cos 4 \theta_{23} \right) \cos 2 \delta \sin \eta \right]\,.
\end{eqnarray}
Actually, the leading term of Eq.~(\ref{eq:DEL_IHAP}),
\ie, $\delta_\nu{}|_{r_m =0}$ is obtained from $d_1 + d_2$ as,
\begin{eqnarray}
  d_1 +d_2 \eqn{=} 
  \frac{1}{4} \sin 2 \theta_{12} \cos^2 \theta_{13} \Big[ \left\{ - 5 - 3 \cos 4 \theta_{23} + \cos 2 \theta_{13} \left( 7 + \cos 4 \theta_{23} \right) \right\} \sin \eta \nonumber\\
  \eqn{}+ 4 \sin \theta_{13} \sin 4 \theta_{23} \left( \sin \delta \cos \eta - \cos 2 \theta_{12} \cos \delta \sin \eta \right) \Big]\,.
\end{eqnarray}
This conclusion is consistent with Eq.~(\ref{eq:DEL_IH}).



\begin{thebibliography}{100}

\bibitem{Riotto:1999yt}
  A.~Riotto and M.~Trodden,
  Ann.\ Rev.\ Nucl.\ Part.\ Sci.\  {\bf 49} (1999) 35
  [arXiv:hep-ph/9901362].

\bibitem{Fukugita:1986hr}
  M.~Fukugita and T.~Yanagida,
  Phys.\ Lett.\  B {\bf 174} (1986) 45 .
  
\bibitem{Buchmuller:2005eh}
  W.~Buchmuller, R.~D.~Peccei and T.~Yanagida,
  Ann.\ Rev.\ Nucl.\ Part.\ Sci.\  {\bf 55} (2005) 311
  [arXiv:hep-ph/0502169].

\bibitem{Seesaw}
P.~Minkowski,
Phys.\ Lett.\ B {\bf 67} (1977) 421;
T.~Yanagida,
in {\em Proc. of the Workshop on the Unified Theory
and the Baryon Number in the Universe}, 
Tsukuba, Japan, Feb.~13-14, 1979, p.~95, 
eds. O.~Sawada and S.~Sugamoto, 
(KEK Report KEK-79-18, 1979, Tsukuba); 
Progr.\ Theor.\ Phys.\ {\bf 64} (1980) 1103 ; 
M.~Gell-Mann, P.~Ramond and R.~Slansky, 
in {\em Supergravity}, 
eds. P.~van~Niewenhuizen and D.~Z.~Freedman
(North Holland, Amsterdam 1980);
P.~Ramond, 
in {\em Talk given at the Sanibel Symposium}, 
Palm Coast, Fla., Feb.~25-Mar.~2, 1979, preprint CALT-68-709
(retroprinted as hep-ph/9809459);
S.~L.~Glashow,
in {\em Proc. of the Carg\'ese  Summer Institute on Quarks and Leptons},
Carg\'ese, July 9-29, 1979, 
eds. M.~L\'evy et. al, , (Plenum, 1980, New York), p707.

  
\bibitem{Giudice:2003jh}
  G.~F.~Giudice, A.~Notari, M.~Raidal, A.~Riotto and A.~Strumia,
  Nucl.\ Phys.\  B {\bf 685} (2004) 89
  [arXiv:hep-ph/0310123].

\bibitem{Asaka:2005pn}
  T.~Asaka and M.~Shaposhnikov,
  Phys.\ Lett.\  B {\bf 620} (2005) 17.

\bibitem{Asaka:2005an}
  T.~Asaka, S.~Blanchet and M.~Shaposhnikov,
  Phys.\ Lett.\  B {\bf 631} (2005) 151.

\bibitem{Gorbunov:2007ak}
  D.~Gorbunov and M.~Shaposhnikov,
  JHEP {\bf 0710} (2007) 015
  [arXiv:0705.1729 [hep-ph]].

\bibitem{Akhmedov:1998qx}
  E.~K.~Akhmedov, V.~A.~Rubakov and A.~Y.~Smirnov,
  Phys.\ Rev.\ Lett.\  {\bf 81} (1998) 1359.

\bibitem{Shaposhnikov:2008pf}
  M.~Shaposhnikov,
  JHEP {\bf 0808} (2008) 008
  [arXiv:0804.4542 [hep-ph]].

\bibitem{Kuzmin:1985mm}
  V.~A.~Kuzmin, V.~A.~Rubakov and M.~E.~Shaposhnikov,
  Phys.\ Lett.\  B {\bf 155} (1985) 36.

\bibitem{Asaka:2006nq}
  T.~Asaka, M.~Laine and M.~Shaposhnikov,
  JHEP {\bf 0701} (2007) 091
  [arXiv:hep-ph/0612182].


\bibitem{Laine:2008pg}
  M.~Laine and M.~Shaposhnikov,
  JCAP {\bf 0806} (2008) 031
  [arXiv:0804.4543 [hep-ph]].

\bibitem{Boyarsky:2009ix}
  A.~Boyarsky, O.~Ruchayskiy and M.~Shaposhnikov,
  Ann.\ Rev.\ Nucl.\ Part.\ Sci.\  {\bf 59} (2009) 191
  [arXiv:0901.0011 [hep-ph]].

\bibitem{Casas:2001sr}
  J.~A.~Casas and A.~Ibarra,
  Nucl.\ Phys.\  B {\bf 618} (2001) 171
  [arXiv:hep-ph/0103065].

\bibitem{Schwetz:2008er}
  For example, see,  
  T.~Schwetz, M.~A.~Tortola and J.~W.~F.~Valle,
  New J.\ Phys.\  {\bf 10} (2008) 113011
  [arXiv:0808.2016 [hep-ph]].

\bibitem{AsakaIshida}
  T.~Asaka and H.~Ishida, in preparation.

\bibitem{Khlebnikov:1988sr}
  S.~Y.~Khlebnikov and M.~E.~Shaposhnikov,
  Nucl.\ Phys.\  B {\bf 308} (1988) 885;
  J.~A.~Harvey and M.~S.~Turner,
  Phys.\ Rev.\  D {\bf 42} (1990) 3344.



\end{thebibliography}
\end{document}